\documentclass[prb,aps,amsmath,twocolumn,amssymb,showpacs,superscriptaddress]{revtex4}

\usepackage{graphicx}
\usepackage{bm}
\usepackage[usenames]{color}

\def \be#1\ee {\begin{equation}#1\end{equation}}
\def \bea#1\eea {\begin{eqnarray}#1\end{eqnarray}}

\def\str{\mathop{\rm str}}
\def\tr{\mathop{\rm tr}}

\def\erf{\mathop{\rm erf}}

\def\diag{\text{diag}}
\newcommand{\corr}[1]{\langle #1\rangle}
\renewcommand{\Re}{\mathop{\rm Re}}
\renewcommand{\Im}{\mathop{\rm Im}}

\def\vp{\varphi}
\def\br{{\bf r}}

\def\ETh{E_{\text{Th}}}

\begin{document}

\title{Superconducting proximity effect in quantum wires without time-reversal symmetry}

\author{M.~A.~Skvortsov}
\affiliation{L.~D.~Landau Institute for Theoretical Physics, 142432 Chernogolovka, Russia}
\affiliation{Moscow Institute of Physics and Technology, 141700 Moscow, Russia}

\author{P.~M.~Ostrovsky}
\affiliation{Max Planck Institute for Solid State Research, Heisenbergstr.\ 1,
70569 Stuttgart, Germany}
\affiliation{L.~D.~Landau Institute for Theoretical Physics, 142432 Chernogolovka, Russia}

\author{D.~A.~Ivanov}
\affiliation{Institute for Theoretical Physics, ETH Z\"urich, 8093 Z\"urich, Switzerland}
\affiliation{Institute for Theoretical Physics, University of Z\"urich, 8057 Z\"urich, Switzerland}

\author{Ya.~V.~Fominov}
\affiliation{L.~D.~Landau Institute for Theoretical Physics, 142432 Chernogolovka, Russia}
\affiliation{Moscow Institute of Physics and Technology, 141700 Moscow, Russia}

\date{February 19, 2013}

\begin{abstract}
We study the superconducting proximity effect in a quantum wire
with broken time-reversal (TR) symmetry connected to a
conventional superconductor. We consider the situation
of a strong TR-symmetry breaking, so that Cooper pairs entering
the wire from the superconductor are immediately destroyed.
Nevertheless, some traces of the proximity effect survive:
for example, the local electronic density of states (LDOS)
is influenced by the proximity to the superconductor,
provided that localization effects are taken into account.
With the help of the supersymmetric sigma model, we calculate the
average LDOS in such a system.
The LDOS in the wire is strongly modified close to the interface
with the superconductor at energies near the Fermi level.
The relevant distances from the interface are of the order
of the localization length, and the size of the energy window
around the Fermi level is of the order of the
mean level spacing at the localization length.
Remarkably, the sign of the effect is sensitive to the way the
TR symmetry is broken: In the spin-symmetric case (orbital magnetic field),
the LDOS is depleted near the Fermi energy, whereas for the broken spin
symmetry (magnetic impurities), the LDOS at the Fermi energy is enhanced.
\end{abstract}


\maketitle

\section{Introduction}

Proximity effect in normal--superconducting structures is a phenomenon of
induced superconducting correlations in the normal metal (N) in an electric
contact with a superconductor (S).\cite{proximity}
Such correlations arise due to
Andreev reflections at the NS interface: an electron from the N part
reflects as a hole by emitting a Cooper pair into the S part.\cite{Andreev} The
extent of such proximity correlations is determined by the structure
and geometry of the N part: in a small N grain, the correlations are uniform
over its volume, while in a large N contact the proximity
correlations extend over some distance determined by the energy of the
electrons relative to the Fermi energy.\cite{ZZ81,BBS96}

One of the signatures of the proximity effect is the modification of the
local electronic density of states (LDOS)
in the N part. Such a modification is
most pronounced in the case of chaotic electron dynamics in the N part,
when a so-called ``minigap'' is formed in the
density of states.\cite{GKI04}
The size of the minigap can be estimated as
$\min(\Delta,1/\tau_\text{esc})$, where
$\Delta$ is the superconducting gap and
$\tau_\text{esc}$
is the time required for an electron in the N region to establish a contact
with the superconductor\cite{Taras-Semchuk-Altland,Ehrenfest-comment}
(hereafter we assume $\hbar=1$).
In particular, for a long diffusive wire (with its length $L$ exceeding
the superconducting coherence length) and transparent NS interface,
the minigap is of the order of the Thouless energy,\cite{ZhouCharlat}
\begin{equation}
  E_{\rm Th} = D/L^2 \, ,
\end{equation}
where $D$ is the diffusion coefficient.

The proximity effect is sensitive to the time-reversal (TR) symmetry
in the N part, which is necessary for superconducting correlations.
If the TR symmetry is broken (e.g., by a strong magnetic field or
by magnetic impurities), then the conventional quasiclassical
theory predicts that
no proximity effect can survive beyond the distance over which
the TR symmetry is broken.\cite{BBS96}

However, this quasiclassical description is known to be
incomplete. The most prominent example of a proximity effect
in the absence of the TR symmetry is the random-matrix theory (RMT):
indeed, in superconducting symmetry classes with broken TR
symmetry, the density of states is
modified in the energy window of the order of the interlevel
spacing around the Fermi energy.\cite{AZ} In extended systems,
the perturbative modes (diffusons and possibly cooperons,
depending on the symmetry) responsible for the proximity effect
beyond the quasiclassical approximation have been identified
in Refs.~\onlinecite{AZ,Taras-AdvPhys}.
The interplay between such mesoscopic fluctuations
and localization effects was studied in various
superconducting and chiral symmetry classes in
Refs.~\onlinecite{gade:91,nersesyan:94:95,senthil:98,Altland-Simons-Zirnbauer:02}.

In our present paper, we consider another example
of a non-quasiclassical superconducting
proximity effect in disordered quantum wires with
broken TR symmetry. Under the assumption of a quantum
coherence, the relevant length and energy scales in such
systems are determined by the Anderson
localization.\cite{anderson:58} The length scale
at which the LDOS is modified due to the proximity
effect is given by the localization length $\xi$.
The corresponding energy scale $\Delta_\xi$
is given by the level spacing between
states at the length $\xi$.
In one-dimensional geometry,\cite{Delta-xi}
\begin{equation}
\label{Delta_xi}
  \Delta_\xi = D/\xi^2 \, .
\end{equation}

\begin{figure}
\includegraphics[width=.4\textwidth]{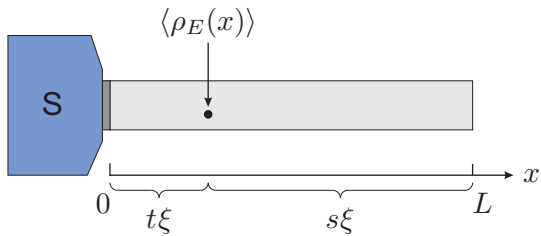}
\caption{(Color online)
Quantum wire of length $L$ coupled to a massive superconductor
at $x=0$. We calculate the average local density of states
$\corr{\rho_E(x)}$ for an arbitrary relation between
$x$, $L$, and the localization length $\xi$.
The dimensionless distances $t$ and $s$ are used
in Eq.~(\ref{full-wave-function-definition}).}
\label{F:setup}
\end{figure}

We illustrate this qualitative picture with an explicit
calculation in the model of a quantum wire with a large
number of conducting channels. Namely, we consider such a wire
(of a finite length $L$) with a broken
TR symmetry in a contact with a superconductor (Fig.~\ref{F:setup}).
Using the method of
the nonlinear supersymmetric sigma-model,\cite{Efetov1983,Efetov-book}
in conjunction with recent exact results for localization
in quasi-one-dimensional unitary wires,\cite{SO07,DM,IOS2009}
we calculate the average LDOS as a function of energy
and coordinate along the wire in different limiting cases.

A remarkable detail of our analysis is the two different ways of
breaking the TR symmetry. Namely, it may be broken either {\it without}\
breaking the spin-rotational symmetry (e.g., by a strong magnetic field
inducing a TR-symmetry-breaking vector potential, but only a negligible
Zeeman field) or {\it with}\ breaking the spin-rotational symmetry
(e.g., by including magnetic impurities coupled to the spins of
electrons). These two possibilities correspond to different symmetry
classes in the RMT classification (C and D, respectively)\cite{AZ}
and exhibit quite different types of the proximity effect. First,
while in class C the LDOS is suppressed at low energies and at short
distances, in class D it is enhanced (similarly to the
RMT results). Second, in long wires ($L \gg \xi$) at small energies
($E \ll \Delta_\xi$), the proximity effect extends to
the Mott length scale\cite{Mott}
\be
\label{LMott}
  L_M = 2\xi \ln(\Delta_\xi/E)
\ee
in class C, while only to the localization length $\xi$ in class D
(Fig.\ \ref{F:rho-E0-x}).
Such a behavior is related to the presence (absence)
of repulsion between an energy level and its mirror counterpart
in the symmetry class C (D) and is discussed in Sec.~\ref{S:Discussion}.

The paper is organized as follows.
In Sec.~\ref{S:Qualitative} we explain
the relation of the proximity effect
in the absence of the TR symmetry to Anderson localization.
Evaluation of the LDOS for a quasi-one-dimensional wire
with the broken TR symmetry is reduced to the unitary sigma-model
in Sec.~\ref{S:Sigma-model}, and its exact solution is presented
in Sec.~\ref{S:Exact}. Resulting expressions for the LDOS in various
limiting cases are derived in Sec.~\ref{S:Results}.
Our findings are summarized in Sec.~\ref{S:Discussion}.
Technical details are relegated to several Appendices.

\section{Proximity effect without time-reversal symmetry: Role~of~localization}

\label{S:Qualitative}

In a diffusive system
(without TR symmetry breaking),
the superconducting proximity effect is described by the
quasiclassical Usadel equation \cite{Usadel}.
For a given energy, $E$, superconducting correlations decay
into the normal region at the diffusive length scale
\be
\label{LE}
  L_E = \sqrt{D/E} \, .
\ee
The TR symmetry is needed to establish particle-hole
correlations (the soft ``cooperon'' modes\cite{AltlandSimonsBook}) in the wire.
In this paper, we consider the opposite situation:
the proximity effect in the case of the broken TR symmetry.
It may be broken in two different ways: by an external magnetic field
(symmetry class C) or by magnetic impurities (symmetry class D).
In both cases, the spectrum of the cooperon modes acquires a gap
leading to their exponential decay at some characteristic scale $L_c$. This
scale is set either by the magnetic length or by the spin diffusion length. As
long as $L_c$ is shorter than $L_E$ (at sufficiently low energies and, e.g.,
strong magnetic field), superconductive coherence brought into the normal
metal by cooperon modes exists only in a thin layer of length $L_c$ near the
boundary, decaying exponentially at larger distances.

This exponential decay of superconductive correlations at $L\gg L_c$
follows from the Usadel equation.\cite{BBS96}
The latter is an effective tool for nonperturbative summation
of tree-like diagrams\cite{SLF2001} but neglects loop corrections
responsible for quantum localization. We will show below that
these corrections, though suppressed as $1/N$
($N \gg 1$ is the number
of the conducting channels in the wire), give rise to the proximity
effect in the absence of the TR symmetry.

Localization length in a N wire with the broken TR symmetry
is given by\cite{Dorokhov,EL83,Efetov-book}
\be
\label{xi-def}
  \xi
  =
  2 \pi \nu A D
  \sim N l \, .
\ee
Here $A$ is the wire cross section, $l$ is the mean-free path,
and $\nu$ is the bulk density of states, whose definition depends
on spin degeneracy:
\begin{itemize}
\item
In class C (an external magnetic field), the spin symmetry
is preserved, and hence all electron states are doubly degenerate.
In this case, the density of states $\nu$ [in Eq.\ (\ref{xi-def})
and throughout the paper] is defined per spin projection.
\item
In class D (magnetic impurities), the disorder mixes spin components,
and $\nu$ is defined as the total density of states including spin.
\end{itemize}

\begin{figure}
\includegraphics{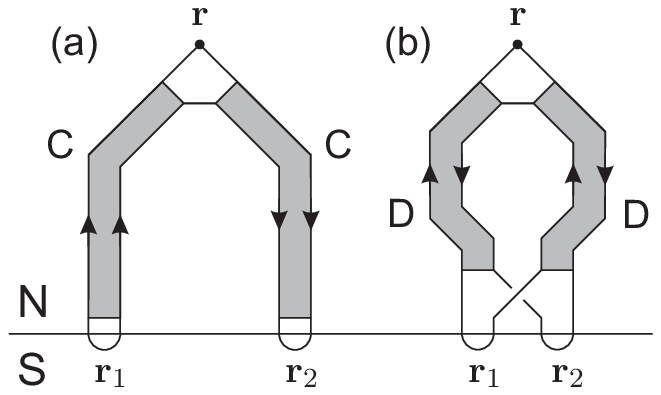}
\caption{
Diagrams for the proximity-induced correction to the local density of
states $\corr{\rho_E(\br)}$ to the lowest-order in the tunneling
transparency of the SN interface:
(a) TR-symmetric case,
when Cooper pairs independently tunnel at $\br_1$ and $\br_2$ and propagate
in the N part as cooperons (shadowed);
(b) broken TR-symmetry case,
when Cooper pairs tunnel at $\br_1\approx\br_2$, propagating further as two
diffusons (shadowed).
}
\label{F:scattering}
\end{figure}

Relation between the proximity effect in the absence of the TR symmetry
and localization can be
visualized
diagrammatically
in the limit of poor transparency of the SN interface
when all transmission coefficients of the barrier are small, $T\ll1$.
Then the influence of the superconductor can be treated perturbatively.
In the TR-symmetric case, the leading process is shown
in Fig.~\ref{F:scattering}(a): A Cooper pair tunnels at a point $\br_1$,
propagates to the observation point $\br$, and returns back to
the superconductor at a point $\br_2$. Integrations over $\br_{1,2}$
taken across the wire section are independent, and the resulting
correction to the LDOS
$\corr{\delta\rho_E(\br)}$ is proportional to $N^2T^2$.
In the case of broken TR symmetry, cooperons are suppressed,
but the LDOS is still affected by the process depicted
in Fig.~\ref{F:scattering}(b):
A tunneling Cooper pair needs to be converted to a pair of diffusons
which can reach the observation point. For such a process,
the tunneling coordinates nearly coincide, $\br_1\approx\br_2$,
and integration over them brings the first power of the wire cross
section: $\corr{\delta\rho_E(\br)} \propto NT^2$.
At the same time we see that the spatial behavior of the LDOS
corrections (a) and (b) are identical: both decay at the scale $L_E$.
The $1/N\propto1/\xi$ suppression of the LDOS correction
in the absence of the TR symmetry is reminiscent of localization
and is related to one-loop structure of the diagram (b),
contrary to the tree-like diagram (a) accounted for
by the Usadel equation.

\section{Reduction to the unitary sigma-model}
\label{S:Sigma-model}

Electronic states in superconducting systems are described
with the help of the Bogolyubov--de Gennes (BdG) Hamiltonian,
which acts as a matrix in the Nambu space:
\be
  \hat{\cal H}_\text{BdG}
  =
  \begin{pmatrix}
    {\cal H} & \Delta \\
    \Delta^* & -\Theta {\cal H}^* \Theta^{-1}
  \end{pmatrix} \, ,
\label{BdG}
\ee
where ${\cal H}$ is the single-particle Hamiltonian, $\Delta$ is the order parameter
field, and $\Theta$ is a unitary matrix that defines the time-reversal
operation: $\psi \mapsto \Theta \psi^*$.\
If the spin symmetry is preserved (class C), we write the BdG Hamiltonian
(\ref{BdG}) for one spin projection (e.g., spin-up
electrons and spin-down holes) and fix $\Theta = 1$.
On the contrary, in class D, when the single-particle spin dynamics
is nontrivial,  we include both spin projections and put $\Theta = is_y$
(the Pauli matrix in the spin space). Thus, in class D the dimension
of the Hamiltonian (\ref{BdG}) is twice larger than in class C.

We define the quasiparticle LDOS normalized to the bulk value as
\be
\label{rho_Green}
  \rho_E(\br) = - \frac{1}{2\pi\nu}
  \Im \tr
  {\cal G}^R_E(\br,\br)\, .
\ee
It is expressed in terms of the retarded Green function of the BdG
Hamiltonian (\ref{BdG}),
\be
  {\cal G}^R_E = (E - \hat{\cal H}_\text{BdG} + i0)^{-1} \, .
\ee
Note that the definition of $\nu$ in Eq.~(\ref{rho_Green}) is
different in the C and D classes (see Section \ref{S:Qualitative}),
which is consistent with the difference in the dimensionality of
$\hat{\cal H}_\text{BdG}$.

In a field-theoretical language,
\cite{SKF98,Bundschuh99,Taras-JETPL,Taras-AdvPhys,OSF01}
disorder averaging of $\rho_E(\br)$ is performed by representing
${\cal G}^R_E$ as a supersymmetric functional integral and following
the standard line of the sigma-model derivation \cite{Efetov-book}.
The details of the derivation are presented
in Appendix \ref{A:D}, and below we summarize the results.

In the normal part ($\Delta=0$) of a hybrid NS system,
the Nambu-Gor'kov Green function ${\cal G}^R_E$ essentially
involves a pair of the retarded and advanced normal-metal
Green functions with opposite energies, $G^R_E$ and $G^A_{-E}$,
which get coupled due to Andreev reflection off the
superconducting order parameter.
Therefore the sigma-model for $\corr{{\cal G}_E^R}$ in the normal region
of an NS system can be exactly rewritten in terms of Evetov's sigma-model
for the product $\corr{G^R_EG^A_{-E}}$.
This relation has been recently demonstrated in Ref.\ \onlinecite{Koziy},
where an explicit mapping between the two models was constructed.
In our problem, $\corr{\rho_E(\br)}$ may be calculated in terms
of the usual Efetov's sigma model written for the normal wire,
but supplemented with a boundary condition at the NS interface
responsible for Andreev reflections.

The TR symmetry in the wire can be broken either by the orbital
magnetic field or by magnetic impurities. In the former case
each level is double degenerate due to the spin symmetry,
whereas in the latter case this Kramers degeneracy is lifted.
For sufficiently strong symmetry breaking \cite{com-crossover}
(or at length scale larger than the length $L_c$ associated
with the TR symmetry breaking)
Models IIa and IIb in Efetov's classification\cite{Efetov-book}
are realized.
In both cases, the cooperon degrees of freedom are frozen out,
and the resulting sigma model is written in terms of
a $4\times4$ supermatrix acting in
the Fermi--Bose (FB) and retarded--advanced (RA) spaces.

For a normal metal, the Models IIa and IIb are mathematically equivalent
(unitary sigma model). This is not the case for the normal part
of a hybrid system, since the form of the effective boundary condition
at the NS interface is sensitive to the way the TR symmetry is broken.
In the presence of Andreev scattering, the Models IIa and IIb correspond
to the symmetry classes C and D in the classification of Ref.~\onlinecite{AZ}.

Thus, in the limit of strongly broken TR symmetry, the proximity effect
in the normal part of an NS system is described by Efetov's
unitary sigma-model. The average (normalized) LDOS
is given by the functional integral over the normal-metal region:
\be
\label{rho-def1}
  \corr{\rho_E(\br)}
  =
  \frac14
  \Re
  \int
  \str (k \Lambda Q(\br))
  e^{-S[Q]-S_\Gamma[Q]} \,
  d[Q]
  \, .
\ee
The action of the model is separated into the bulk diffusive part $S[Q]$
and the boundary term $S_\Gamma[Q]$, derived in Appendix \ref{A:D}. The bulk
action has the standard form for the unitary class sigma model,
\be
\label{SD4}
  S
  =
  \frac{\pi\nu}{4}
  \int d\br \,
  \str
  \left[
    D (\nabla Q)^2
  + 4iE \Lambda Q
  \right] \, ,
\ee
with the $4\times4$ supermatrix $Q$ acting in the FB and RA spaces.
The matrix $\Lambda=\sigma_z^\text{RA}$ is the metallic saddle point,
and the supersymmetry breaking matrix $k=\sigma_z^\text{FB}$
(we follow notations of Ref.~\onlinecite{Efetov-book}).
As before, $D$ is the diffusion coefficient
and $\nu$ is the density of states at the Fermi level
per one spin projection if the spin is conserved (class C) and
including both spin projections
in the case of broken spin symmetry (class D).

The boundary action is derived in Appendix \ref{A:D}
(see also Ref.~\onlinecite{Koziy}).
Throughout the paper, we assume for simplicity
that the superconducting gap $\Delta$ in the S part is large.
The precise condition on $\Delta$ is formulated in Eq.~(\ref{large-Delta}).
Under this assumption, the boundary action simplifies to the form
\be
\label{Sboundary4}
  S_\Gamma = - \frac{1}{2} \sum_i
  \str \ln [1-e^{-4\beta_i} Q(0) \Xi Q^T(0) \Xi^T] \, .
\ee
Here the parameters $\beta_i$ are related
to the transmission coefficients at the SN interface:
$T_i=1/\cosh^2\beta_i$, with $i$ labeling open channels.
In the case of conserved spin (class C), the channels are double degenerate
but the index $i$ counts them only once.
With broken spin symmetry (class D), this degeneracy is lifted.
In both cases, the normal-state conductance can be written
as a product $G_0 g_N$, where $g_N = \sum_i T_i$, and the conductance
quantum is defined as $G_0=2e^2/h$ for class C and $G_0=e^2/h$ for class D.

The main ingredient which distinguishes between the symmetry
classes C and D is the matrix $\Xi$ entering Eq.~(\ref{Sboundary4}).
For class C it has been derived in Ref.\ \onlinecite{Koziy},
and derivation for both classes is presented in Appendix~\ref{A:D}:
\begin{subequations}
\label{Xi}
\begin{align}
\label{Xi-C}
  \text{class C:}
  \qquad
  \Xi
  =
  \begin{pmatrix}
    i\sigma_y^\text{RA} & 0 \\
    0 & \sigma_x^\text{RA}
  \end{pmatrix}_\text{FB}
  \, ,
\\[3pt]
\label{Xi-D}
  \text{class D:}
  \qquad
  \Xi
  =
  \begin{pmatrix}
    \sigma_x^\text{RA} & 0 \\
    0 & i\sigma_y^\text{RA}
  \end{pmatrix}_\text{FB}
  \, .
\end{align}
\end{subequations}
The matrix $\Xi$ has a nontrivial structure
in the superspace, which is the mathematical reason why
Eq.~(\ref{rho-def1}) results in a nontrivial LDOS.

Note that the matrix $\Xi$ does not belong to the standard
unitary manifold for $Q$ matrices:
it does not obey the condition $\Xi^2=1$ [in the F (B) sector for class
C (D), respectively].
One finds that the
saddle-point solution for the action $S[Q]+S_\Gamma[Q]$
is simply $Q(\br)=\Lambda$, leading to the metallic density of states, $\rho(\br)=1$,
indicating no proximity effect at the level of the Usadel equation.
However proximity effect absent in the quasiclassical approximation will
manifest itself once fluctuations around $\Lambda$ are taken into account.
The same fluctuations are responsible for localization.
Thus the superconducting proximity effect in a normal metal with
broken TR symmetry is inevitably related to localization,
leading to strong modification of the LDOS near the SN interface
at energies $E\lesssim\Delta_\xi$ and length scales $x\lesssim\xi$.

\section{Analytic solution for quasi-1D wires}
\label{S:Exact}

Here we apply the general framework described in the previous Section
in order to calculate the average LDOS, $\corr{\rho_E(x)}$,
in a finite quasi-one-dimensional wire of length $L$ coupled to a superconductor at $x=0$,
see Fig.~\ref{F:setup}.
We will work in the
limit of large $\Delta$, which is spelled out in Eq.~(\ref{large-Delta})
of Appendix \ref{A:D},
and assume that the TR symmetry in the wire is completely broken.

The one-dimensional sigma model (\ref{rho-def1}) can be solved
exactly by mapping onto effective quantum mechanics,\cite{EL83}
with the $x$ coordinate playing the role of the imaginary time.
In this formalism, evaluation of the functional integral (\ref{rho-def1})
is reduced to solving Schr\"odinger equations
for the wave functions $\Psi(Q)$ on the sigma-model manifold.
This technical procedure described in Appendix \ref{A:D}
leaves us with the object $\Psi(\lambda_F,\lambda_B)$ depending only
on the ``eigenvalues'' $\lambda_F$ and $\lambda_B$ of the $Q$ matrix.

The main ingredient for calculation of $\corr{\rho_E(x)}$ is the wave
function $\Psi(\lambda_F,\lambda_B;t,s)$ which accounts for coherent motion
of the electron and hole in the wire. It can be obtained by successive
application of two evolution operators on unity (corresponds to open
boundary conditions at the free end of the wire):
\be
  \Psi(\lambda_F,\lambda_B;t,s)
  =
  e^{-2\tilde Ht} e^{-2Hs} \circ 1 \, ,
\label{full-wave-function-definition}
\ee
where (see Fig.~\ref{F:setup})
\be
  t=x/\xi,
\quad
  s=(L-x)/\xi \, .
\ee
The Hamiltonians $\tilde H$ and $H$ govern imaginary-time evolution
at the segments $[0,x]$ and $[x,L]$, respectively.
They have the form \cite{EL83}
\begin{multline}
  H = -\frac{(\lambda_B-\lambda_F)^2}2
  \biggl[
    \frac{\partial}{\partial\lambda_F}
    \frac{1-\lambda_F^2}{(\lambda_B-\lambda_F)^2}
    \frac{\partial}{\partial\lambda_F}
\\ {}
  +
    \frac{\partial}{\partial\lambda_B}
    \frac{\lambda_B^2-1}{(\lambda_B-\lambda_F)^2}
    \frac{\partial}{\partial\lambda_B}
  \biggr]
  + \frac{\kappa^2}{16} (\lambda_B-\lambda_F) \, ,
\label{H}
\end{multline}
and \cite{DM,IOS2009}
\be
\label{tilde-H}
  \tilde H
  = (\lambda_B-\lambda_F)^{-1} H (\lambda_B-\lambda_F)
  = \tilde H_B + \tilde H_F
\ee
with
\begin{subequations}
\begin{align}
\label{tilde-HB}
  & \tilde H_B
  = - \frac12 \frac{\partial}{\partial\lambda_B} (\lambda_B^2-1)
\frac{\partial}{\partial\lambda_B}
  + \frac{\kappa^2}{16} \lambda_B
  \, ,
\\
\label{tilde-HF}
  & \tilde H_F
  = - \frac12 \frac{\partial}{\partial\lambda_F} (1-\lambda_F^2)
\frac{\partial}{\partial\lambda_F}
  - \frac{\kappa^2}{16} \lambda_F \, ,
\end{align}
\end{subequations}
where the dimensionless quantity $\kappa$ stands for
\be
  \kappa^2
  =
  - \frac{8iE}{\Delta_\xi}
\label{kappa-def}
\ee
[it is consistent with the notations of Refs.~\onlinecite{SO07}
and \onlinecite{IOS2009}, with the frequency $\omega=E-(-E)=2E$].

Unfortunately, the function $\Psi(\lambda_F,\lambda_B;t,s)$ cannot
be generally obtained in a closed form. The only exception is the
zero mode of the Hamiltonian $H$,
$\Psi_0(\lambda_F,\lambda_B) \equiv \Psi(\lambda_F,\lambda_B;0,\infty)$,
given by \cite{SO07}
\be
\label{Psi0}
  \Psi_0(\lambda_F,\lambda_B)
  =
  I_0(q) p K_1(p) + q I_1(q) K_0(p) \, ,
\ee
where
\be
\label{pq}
  p = \kappa \sqrt{(\lambda_B+1)/2} \, ,
\qquad
  q = \kappa \sqrt{(\lambda_F+1)/2}
\ee
and $I_n$ and $K_n$ are the modified Bessel functions.

The function $\Psi(\lambda_F,\lambda_B;t,s)$ should be finally integrated
over $Q(0)$ with the weight $e^{-S_\Gamma[Q]}$ in order to obtain
$\corr{\rho_E(x)}$. At this stage calculations for the symmetry
classes C and D are different due to a different form of the
superconducting matrix (\ref{Xi}).

\subsection{General expression for class C}

For the symmetry class C, calculations presented in Appendix
\ref{A:derivation} lead to the following exact expression:
\begin{multline}
\label{rho(x)C}
  \corr{\rho_E(x)}
  =
  1
  +
  \frac12 \Re
  \int_{-1}^1
  d\lambda_F
  \int_1^\infty
  d\lambda_B
\\ {}
  \times
  \frac{\partial e^{-S_0^\text{C}(\lambda_B)}}{\partial\lambda_B} \,
  \Psi(\lambda_F,\lambda_B;t,s) \, .
\end{multline}
Parameters of the SN interface are encoded in the boundary action:
\be
\label{SGamma0C}
  S_0^\text{C}(\lambda_B)
  =
  \frac12
  \sum_i
  \ln
  \left[
    1 + {\cal T}_i \, (\lambda_B^2-1)
  \right]
  \, ,
\ee
where
\begin{equation}
{\cal T}_i = \frac{T_i^2}{(2-T_i)^2}
\label{Andreev-transmission}
\end{equation}
is the Andreev transmission of the $i$'th channel.\cite{Andreev-transmission}

The boundary action (\ref{SGamma0C}) suppresses fluctuation
of the bosonic variable $\lambda_B$.
The strength of the coupling to the superconductor is characterized
by the dimensionless [in units of $G_0$ defined below Eq.~(\ref{Sboundary4})]
Andreev conductance of the interface,
\begin{equation}
\label{GA}
  g_A = 2 \sum_i \, {\cal T}_i \, .
\end{equation}
In the limit of $g_A\gg1$, the integral over $\lambda_B$
in Eq.~(\ref{rho(x)C}) comes from $\lambda_B-1\lesssim1/g_A$.
If $\Psi(\lambda_F,\lambda_B;t,s)$ is a slow function
of $\lambda_B$ in this region then Eq.~(\ref{rho(x)C})
simplifies to
\be
\label{rho(x)-large-GA}
  \corr{\rho_E(x)}
  =
  1
  -
  \frac{1}{2}
  \Re
  \int_{-1}^1
  d\lambda_F
  \Psi(\lambda_F,1;t,s) \, .
\ee
As we will see below, approximation (\ref{rho(x)-large-GA}) applies if
\be
\label{GA-large}
  g_A \gg 1 \quad
  \textrm{and} \quad
  g_A \gg  \min(\xi/L_E,E/\delta) \, ,
\ee
where $\delta=(2\nu AL)^{-1}$ is the mean level spacing in the wire.

\subsection{General expression for class D}

In close analogy with Eq.~(\ref{rho(x)C}), calculations
in Appendix \ref{A:derivation} lead for the symmetry class D:
\begin{multline}
\label{rho(x)D}
  \corr{\rho_E(x)}
  =
  1
  +
  \frac12 \Re
  \int_{-1}^1
  d\lambda_F
  \int_1^\infty
  d\lambda_B
\\ {} \times
  \frac{\partial e^{-S_0^\text{D}(\lambda_F)}}{\partial\lambda_F} \,
  \Psi(\lambda_F,\lambda_B;t,s) \, ,
\end{multline}
with the boundary action
\be
\label{SGamma0D}
  S_0^\text{D}(\lambda_F)
  =
  -
  \frac{1}{2}
  \sum_i
  \ln
  \left[
    1 - {\cal T}_i \, (1-\lambda_F^2)
  \right]
  \, .
\ee
For large $g_A$ [Eq.~(\ref{GA-large})],
$\lambda_F$ is pinned to $\pm1$ and we obtain
\begin{multline}
\label{rho(x)-large-GA-paramag}
  \corr{\rho_E(x)}
  =
  1
  +
  \frac{1}{2}
  \Re
  \int_1^\infty
  d\lambda_B
  \bigl[
  \Psi(1,\lambda_B;t,s)
\\ {}
  -
  \Psi(-1,\lambda_B;t,s)
  \bigr]\, .
\end{multline}

\begin{figure}
\includegraphics{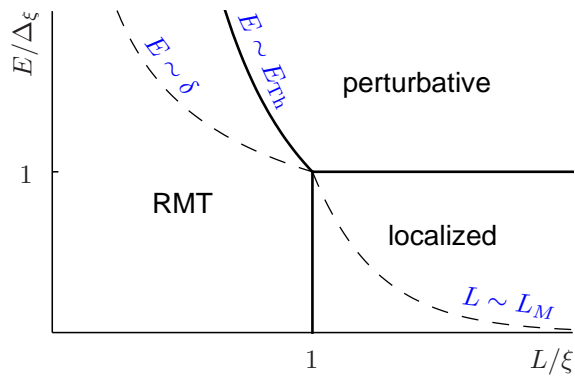}
\caption{(Color online)
Regions of different behavior of the average LDOS
in the coordinates $L/\xi$, $E/\Delta_\xi$.}
\label{F:diagram}
\end{figure}

\section{Results}
\label{S:Results}

Here we analyze the general expressions (\ref{rho(x)C})
and (\ref{rho(x)D}) in various regions of the system parameters,
schematically shown in Fig.~\ref{F:diagram}. We will be mainly
interested in the limit $g_A\to\infty$ [Eq.~(\ref{GA-large})]
when the LDOS is given
by Eqs.~(\ref{rho(x)-large-GA}) and (\ref{rho(x)-large-GA-paramag}),
and present the result for an arbitrary $g_A$ only in the perturbative
regime discussed in Sec.~\ref{SS:pert}.

\subsection{Random-matrix regime ($L\ll\xi$ and $E\ll\ETh$)}

\begin{figure}
\includegraphics{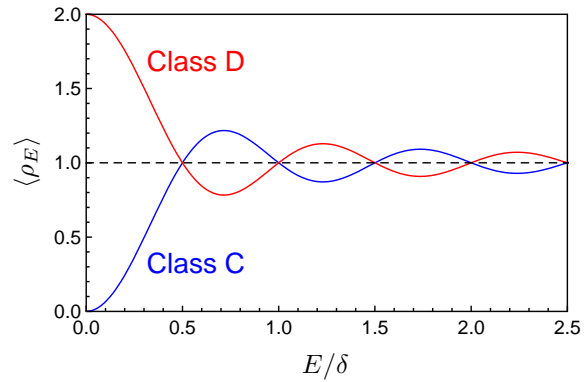}
\caption{(Color online)
Energy dependence of the RMT density of states
for the symmetry classes C and D.}
\label{F:RMT}
\end{figure}

We start the analysis of the general expression (\ref{rho(x)-large-GA})
for the LDOS with the simplest case of short wires, $L\ll\xi$.
In the limit of $E\ll\ETh$, electrons have enough time to explore
the whole available space and one should recover the RMT statistics.
In this case, one can neglect the terms with derivatives in the
Hamiltonians: $H \approx \tilde H \approx (\kappa^2/16)(\lambda_B-\lambda_F)$,
and one simply gets
\be
\label{Psi(t,s)-largeE}
  \Psi(\lambda_F,\lambda_B;t,s)
  =
  e^{-(\kappa^2/8)(\lambda_B-\lambda_F)(L/\xi)} \, ,
\ee
independently of $x$. Substituting into Eqs.~(\ref{rho(x)-large-GA})
and (\ref{rho(x)-large-GA-paramag}) one readily recovers
the RMT results for the symmetry classes C (minus sign)
and D (plus sign):\cite{AZ}
\be
\label{RMT}
  \corr{\rho_E(x)}
  =
    1 \mp \frac{\sin(2\pi E/\delta)}{(2\pi E/\delta)}
  \, ,
\ee
with the mean level spacing
\begin{equation}
\delta=(2\nu AL)^{-1}
\label{delta-definition}
\end{equation}
[this definition of the level spacing takes into account
both electron and hole states, hence it contains an additional
factor $1/2$].
Energy dependence of the RMT density of states is shown in Fig.~\ref{F:RMT}.

The gradient terms in the sigma model
may be taken into account perturbatively, which results in
a position-dependent correction to the RMT density of states (\ref{RMT}):
\begin{multline}
 \corr{\delta\rho_E(x)}
  = \pm \frac{L}{\xi} \bigg\{
      \left[ \frac{1}{3} - \left( 1 - \frac{x}{L} \right)^2 \right]
        \left( 1 \mp \frac{\sin (2\pi E/\delta)}{2\pi E/\delta} \right) \\
      +\frac{1 \mp \cos (2\pi E/\delta)}{3} + O(L/\xi) + O(E/E_\text{Th})
    \bigg\}\, ,
\label{rho-RMT-correction}
\end{multline}
with the upper (lower) signs corresponding to the symmetry class
C (D). Note that in the symmetry class C both the main result (\ref{RMT})
and the correction (\ref{rho-RMT-correction}) vanish at zero energy.
The gradient terms in the Hamiltonians $H$ and $\tilde H$ become significant
if the length of the wire $L$ exceeds $\xi$ or at high energies
$E \gtrsim \ETh$. This establishes the boundaries of the RMT regime,
see Fig.\ \ref{F:diagram}.

\subsection{Perturbative regime ($E\gg\ETh,\Delta_\xi$)}
\label{SS:pert}

For $E\gg \max (\ETh,\Delta_\xi)$, the LDOS is close to 1 everywhere
in the wire, with the difference $\corr{\rho_E(x)}-1$ decaying
at the length scale $L_E$ from the SN boundary.
In this case only small deviations of $\lambda_F$ and $\lambda_B$
from 1 are important. Then the Hamiltonians $H$ and $\tilde H$
can be simplified in the vicinity of $\lambda_F=\lambda_B=1$
(``north pole'' of the fermionic sphere) yielding an effective
linear oscillator model. In this approximation, the wave function
(\ref{full-wave-function-definition}) takes the following form:
\begin{multline}
  \Psi(\lambda_F,\lambda_B; t, s)
  = \frac{\cosh^2 (\kappa s/2)}{\cosh^2 (\kappa (t+s)/2) }
\\{}
  \times
  \exp\left(
      -\frac{\kappa(\lambda_B-\lambda_F)}{4}\, \tanh\frac{\kappa(t+s)}{2}
    \right)\, .
\label{psi-perturbative}
\end{multline}
The density of states for the two symmetry classes then follows
from Eqs.~(\ref{rho(x)C}) and (\ref{rho(x)D}),
where only small deviations of $\lambda_F$ and $\lambda_B$ from 1
are important. In this limit, the boundary actions (\ref{SGamma0C})
and (\ref{SGamma0D}) take the form
$S_0^\text{C}(\lambda_B) = (g_A/2)(\lambda_B-1)$
and
$S_0^\text{D}(\lambda_F) = (g_A/2)(1-\lambda_F)$.
Integrating over $1-\lambda_F$ and $\lambda_B-1$, we obtain
\begin{multline}
\label{rho-pert}
  \corr{\rho_E(x)}
  =
  1 \mp 4 \Re \frac{\cosh^2 \kappa s/2}{\kappa \sinh \kappa (t+s)}
\\{}
  \times
  \frac{g_A}{g_A + (\kappa/2)\tanh(\kappa(t+s)/2)}
  \, .
\end{multline}
As in Eq.~(\ref{RMT}), the minus (plus) sign corresponds to the symmetry
class C (D).

Modification of the metallic LDOS by the proximity effect is maximal
for good SN contacts with $g_A\gg g_N(L_E)$,
where $g_N(L_E) = \xi/L_E = \sqrt{E/\Delta_\xi}$
is the normal-state conductance of the wire of length $L_E$.
In this case, the second factor in the correction (\ref{rho-pert}) may
be approximated by one, and the result (in the physical units) reads
\be
\corr{\rho_E(x)} =
  1 \mp
  \frac{2L_E}{\xi} \Re \frac{\cosh^2 [(1-i)(L-x)/L_E]}{(1-i) \sinh [2(1-i) L /L_E]} \, .
\label{rho-pert-2}
\ee
In the particular case of an infinite wire,
Eq.~(\ref{rho-pert-2}) simplifies to
\be
  \corr{\rho_E(x)} = 1 \mp \frac{L_E}{\sqrt{2}\, \xi}\, e^{-2x/L_E}
  \cos \left(\frac{2x}{L_E} + \frac{\pi}{4} \right)\, .
\label{rho-largeLlargeE}
\ee
[Similar simplifications are also possible in the
opposite limit $g_A\ll g_N(L_E)$: in that case, the
perturbative corrections in Eqs.\ (\ref{rho-pert-2})
and (\ref{rho-largeLlargeE}) get multiplied by $g_A$
and correspond to the diagram in Fig.~\ref{F:scattering}b.]
This result agrees with our argument in
Sec.~\ref{S:Qualitative}: the proximity effect extends
to the normal region at the length scale $L_E$ (just like in
the conventional Usadel equation), but its magnitude is small
as $L_E/\xi\ll1$ (in the perturbative regime).

Note that our perturbative calculation gives only the contribution to
the LDOS from the vicinity of the north pole of the fermionic sphere
($\lambda_F=\lambda_B=1$).
In fact, the expressions (\ref{psi-perturbative})--(\ref{rho-pert-2})
provide a valid perturbative treatment of the north-pole contribution at
energies $E\gg \max (\delta,\Delta_\xi)$. In particular, in the window
$\delta \ll E \ll \ETh$,
the north-pole contribution (\ref{rho-pert-2}) reproduces the
non-oscillating parts of the RMT result (\ref{RMT}),
(\ref{rho-RMT-correction}).

The south pole of the fermionic sphere ($\lambda_F=-1$, $\lambda_B=1$)
is an alternative saddle point of the action, which is responsible for
the terms oscillating in energy with the period $\delta$ in the RMT
[see, e.g., Eqs.\ (\ref{RMT}) and (\ref{rho-RMT-correction})].
\cite{AndreevAltshuler} At energies $E\gg \ETh$, its contribution is
exponentially suppressed, and therefore it was neglected in our
perturbative calculation. However, at energies $E \lesssim \ETh$, the
south-pole contribution exceeds, by absolute value, the correction to
unity in Eq.~(\ref{rho-pert-2}). Note that at $E\gg \max (\delta,
\Delta_\xi)$ the south-pole contribution may also be found in a
perturbative scheme analogous to the one above, but expanding around the
south pole instead. We do not report this calculation here, but only
remark that, within the window $\delta \ll E \ll \ETh$, it can reproduce
the oscillating terms in the RMT expansion (\ref{RMT}) and
(\ref{rho-RMT-correction}). In other words, this energy window
represents an overlap between the RMT and perturbative approaches (if
the south pole is taken into account).

\begin{figure}
\includegraphics{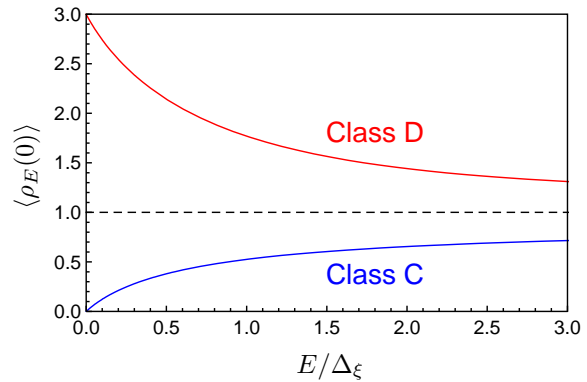}
\caption{(Color online)
Energy dependence of the LDOS in the vicinity of the NS boundary,
$\corr{\rho_E^{(\infty)}(0)}$, for the semi-infinite wire.
The lower (upper) curve corresponds to the symmetry class C (D).}
\label{F:rho0}
\end{figure}

\subsection{Localized regime ($L\gg\xi$ and $E\ll\Delta_\xi$)}

Finally, we turn to the limit of long wires, $L\gg\xi$.
Here the difference between the symmetry classes C and D
becomes more pronounced than just the sign in Eqs.~(\ref{RMT})
and (\ref{rho-pert}), and they require separate considerations.
As in subsection A, our results in this section
are restricted to the ``good contact'' limit
$g_A\gg1$ [see Eq.~(\ref{GA-large})].

\subsubsection{Symmetry class C}

Using Eq.~(\ref{Psi0}) for the zero mode,
one immediately obtains an exact expression for the
energy dependence of the LDOS
at the SN interface for the semi-infinite wire:
\be
\label{rho(0)}
  \corr{\rho_E(0)}
  =
  1
  -
  2 \Re
  \left[
    I_1(\kappa) K_1(\kappa)
  + I_2(\kappa) K_0(\kappa)
  \right] \, ,
\ee
where the imaginary parameter $\kappa^2$ is defined in Eq.~(\ref{kappa-def}).
The plot of this function is shown in Fig.~\ref{F:rho0} (lower curve).
Its asymptotics are
\be
\label{rho(0)-asymp}
  \corr{\rho_E(0)}
  \approx
  \begin{cases}
    \pi E/2\Delta_\xi, & E\ll\Delta_\xi \, , \\[3pt]
    1 - \sqrt{\Delta_\xi/4E}, & E\gg\Delta_\xi \, .
  \end{cases}
\ee
[The $E\gg \Delta_\xi$ asymptotics agrees with
Eq.~(\ref{rho-largeLlargeE}) obtained by perturbative
methods.]

For small $x\ll\xi$, the function $\corr{\rho_E(x)}$ can be obtained
as a power series in $x$ by expanding the evolution operator
$e^{-2\tilde Ht}$ in $t=x/\xi$
(cf.\ Sec.~VI of Ref.~\onlinecite{IOS2009}).
In the nonperturbative limit of $E\ll\Delta_\xi$, this procedure yields
\be
\label{drho/dt-loc}
  \corr{\rho_E(x)}
  =
  \frac{\pi E}{2\Delta_\xi}
  \left(
    1
    +
    2 \frac{x}{\xi}
    +
    2 \frac{x^2}{\xi^2}
    +
    \dots
  \right) \, .
\ee

\begin{figure}
\centerline{
\includegraphics{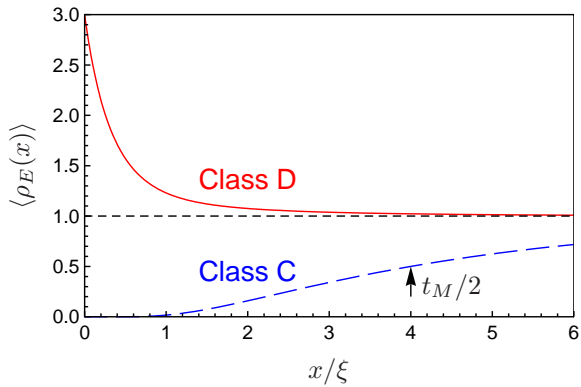}
}
\caption{(Color online)
Dependence of the low-energy
($E\ll\Delta_\xi$) LDOS,
$\corr{\rho_E^{(\infty)}(x)}$, on the distance from the SN interface
for the semi-infinite wire.
In the symmetry class C (lower curve, Eq.~(\ref{rho0(x)-Mott}), sketch),
LDOS depletion extends to half of the Mott scale (\ref{LMott}),
while in the symmetry class D (upper curve, Eq.~(\ref{rho0(x)-D})),
excess LDOS accumulated near the interface
relaxes at the localization length.}
\label{F:rho-E0-x}
\end{figure}

According to Eq.~(\ref{drho/dt-loc}), for small energies, $E\ll\Delta_\xi$,
the LDOS is strongly suppressed
even at the distances from the SN
interface comparable to the localization length, $x\sim\xi$.
In this nonperturbative regime, the LDOS depletion propagates
to a much larger distance of the order of the Mott scale
(\ref{LMott}).
Indeed, in the limit $E\ll\Delta_\xi$, the wave function $\Psi(Q)$
is uniformly spread over the Fermionic sphere,
and Eq.~(\ref{rho(x)-large-GA}) yields
$\corr{\rho_E(x)} = 1 - \Psi(1,1;t,s)$.
To the leading approximation,
$\Psi$ at the north pole
is calculated in Appendix \ref{A:diff-drift},
and one gets (for $|t-t_M/2|\ll t_M$):
\be
\label{rho0(x)-Mott}
  \corr{\rho_E(x)}
  \approx
  \frac12
  +
  \frac12
  \erf
  \left(
    \frac{t-t_M/2}{2\sqrt{t}}
  \right) \, ,
\ee
where $t_M=L_M/\xi$.
This dependence is shown in Fig.~\ref{F:rho-E0-x} (lower curve).

\subsubsection{Symmetry class D}

Substituting the zero mode (\ref{Psi0})
into Eq.~(\ref{rho(x)-large-GA-paramag}),
one readily obtains the energy dependence of the LDOS at the SN interface
for the semi-infinite wire:
\be
\label{rho(0)-paramag}
  \corr{\rho_E(0)}
  =
  1
  +
  2 \Re
  \left[
    I_1(\kappa) K_1(\kappa)
  + \left( I_0(\kappa) - 1 \right) K_2(\kappa)
  \right] \, ,
\ee
with the asymptotic behavior
\be
\label{rho(0)-asymp-paramag}
  \corr{\rho_E(0)}
  \approx
  \begin{cases}
    3 - \pi E/\Delta_\xi, & E\ll\Delta_\xi , \\[3pt]
    1 + \sqrt{\Delta_\xi/4E}, & E\gg\Delta_\xi \, .
  \end{cases}
\ee
[As in class C, the $E\gg \Delta_\xi$ asymptotics agrees with
the perturbative result (\ref{rho-largeLlargeE}).]

The function $\corr{\rho_E(0)}$ is shown in Fig.~\ref{F:rho0}
(upper curve). Remarkably, at the
lowest energies, $E\ll\Delta_\xi$, the LDOS
at the SN boundary is enhanced by the factor of three
compared to the normal case. This behavior should be contrasted
with the LDOS behavior in class C, where
the LDOS near the SN interface is depleted at low energies.

At zero energy, $E=0$, the LDOS can be obtained analytically
for an arbitrary relation between $x$, $L$ and $\xi$.
In the limit $\kappa\to0$, the Hamiltonians $H$ and $\tilde H$
get simplified, which allows us
to evaluate the operator exponents
in Eq.~(\ref{full-wave-function-definition}) with the help of
the Lebedev-Kontorovich transformation \cite{Lebedev-Kontorovich},
see Appendix \ref{A:rho0(x)-D}. The general dependence
of $\corr{\rho_0(x)}$ on the
two parameters $t$ and $s$
given by Eq.~(\ref{rho0-gen}) is quite complicated,
and we present here its limits at $t=0$
and at $s=\infty$.

For the semi-infinite wire, the zero-energy LDOS is given by
\begin{multline}
\label{rho0(x)-D}
  \corr{\rho_0^{(\infty)}(x)}
  =
  1
  -
  \frac{\pi}{4}
  \frac{\partial}{\partial t}
  \biggl\{
    \left(
      1 + e^{-2t}
    \right)
\\ {}
  \times
    \int_0^\infty
      k \, dk \,
      \frac{\sinh(\pi k/2)}{\cosh^2(\pi k/2)}
      e^{-(1+k^2)t/4}
  \biggr\}
  \, ,
\end{multline}
where $t=x/\xi$.
The function $\rho_0^{(\infty)}(x)$ is shown in Fig.~\ref{F:rho-E0-x}.
Near the SN interface, the LDOS exceeds the unperturbed value by
the factor of three.

The excess LDOS accumulated near the boundary with the superconductor
is relaxed to the metallic value at the scale of the localization length
[whereas the LDOS depletion in the C-class case is restored
at a larger Mott length scale, see Eq.~(\ref{rho0(x)-Mott})].

\begin{figure}
\centerline{
\includegraphics{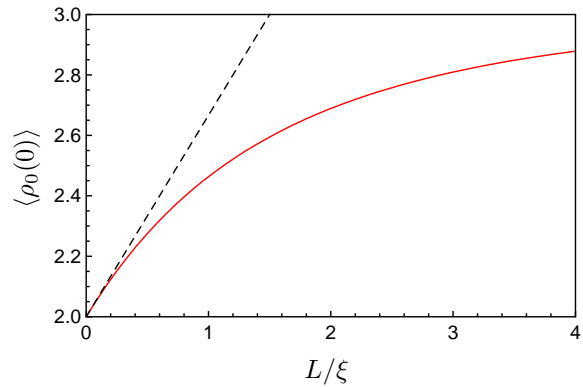}
}
\caption{(Color online)
The zero-energy LDOS at the SN interface for a finite wire
of length $L$ in the symmetry class~D.
The dashed line shows the linear behavior for small $L/\xi$:
$\corr{\rho_0^{(L)}(0)} = 2 + 2L/3\xi + \dots$, see Eq.\
(\ref{rho-RMT-correction}).}
\label{F:rho-0-D-finite}
\end{figure}

For a finite wire of length $L$,
the zero-energy LDOS at the SN interface is given by
\begin{multline}
\label{rho0(0)-D}
  \corr{\rho_0^{(L)}(0)}
  =
  3
  +
  \frac14
  \int_0^\infty
  k\, dk \,
  \tanh\frac{\pi k}{2}
\\{}
  \times
  \biggl[
  \frac{k^2-7}{k^2+1} e^{-(k^2+1)s/4}
  -
  \frac{k^2+1}{k^2+9} e^{-(k^2+9)s/4}
  \biggr]
  \, ,
\end{multline}
where $s=L/\xi$.
This function shown in Fig.~\ref{F:rho-0-D-finite}
interpolates between the RMT value of 2 for short wires ($L\ll\xi$)
and the value of 3 in the localized regime ($L\gg\xi$).
For short wires,
$\corr{\rho_0^{(L)}(0)} = 2 + 2L/3\xi + \dots$
(dashed line in Fig.~\ref{F:rho-0-D-finite}) agrees with
the perturbative result (\ref{rho-RMT-correction}).

\section{Discussion and conclusions}
\label{S:Discussion}

In this paper, we have solved the problem of the LDOS in a
one-dimensional wire without TR symmetry connected to
a massive superconductor. Our calculation confirms qualitative
expectations that the LDOS in the wire is modified in the energy
window of $\Delta_\xi$ around the Fermi level within the distance
of the order of $\xi$ (up to logarithmic corrections)
from the interface with the superconductor.

Propagation of superconductive correlations in the N region
with the broken TR symmetry has been studied previously
in the context of the Josephson coupling in a strong magnetic
field.\cite{SpivakZhou, GalLar01}
It has been shown that while the average Josephson coupling
between a pair of S islands through the N region without the
TR symmetry
decays exponentially, its variance decays only algebraically.
This phenomenon is similar to the universal conductance
fluctuations, \cite{UCF-1,UCF-2} and originates from the
diffuson-only diagram which are insensitive to the TR symmetry breaking.
In the present paper we demonstrate that proximity effect
without the TR symmetry manifests itself already in the average
properties, provided that localization is taken into account.

The observed modification of the LDOS, in the
case of a wire much longer than the localization length $\xi$, is
totally due to localization effects and is absent in the quasiclassical
description (see discussion in Section \ref{S:Qualitative}).
In other words, localization {\it enhances}\/ proximity effect
near the NS interface.
Qualitatively, the effect of localization
is similar to cutting the wire at the distance $\xi$ from the
interface, so that the density of states near the interface becomes
similar to the RMT result in a grain of size $\xi$. While
this similarity is only qualitative in long wires, we indeed find
a true quantitative crossover to the RMT regime in the limit of
short wires (much shorter than $\xi$). Note, however, that both
in the long-wire and short-wire (RMT) limits, the
number of energy levels involved in the modification of the LDOS
is of order one (since $\nu \Delta_\xi \xi A \sim 1$). Thus the
physics of this modification may be understood in terms
of the repulsion of the lowest levels from their own mirror
images (generated by the Andreev reflection), in the spirit of
the superconducting ensembles in the RMT.\cite{AZ}

This picture is in agreement with another important observation:
the sign of the correction to the LDOS depends on the symmetry of
the TR-breaking terms. There are two possible ways to break the
TR symmetry: either conserving the spin symmetry (class C
realizable, e.g., by an orbital magnetic field) or breaking it
(class D realizable, e.g., by magnetic impurities). It is known from
the RMT that corrections to the LDOS at low energies have opposite signs
in these two symmetry classes (see Fig.\ \ref{F:RMT}).
In class C, the levels repel from their own mirror images,\cite{AZ}
and therefore there is an effective repulsion from the Fermi level
leading to a suppression of the LDOS around it. On the other hand,
in class D, there is {\it no repulsion}\/ of any level from its
mirror image,\cite{AZ} and therefore the LDOS is {\it enhanced}\/ around
the Fermi level (levels are ``squeezed'' towards it). According to
the results of our calculations, the same tendency persists also
for long wires, away from the RMT limit
(see Figs.\ \ref{F:rho0} and \ref{F:rho-E0-x}). The above
argument about the repulsion of levels from their mirror images
suggests that our results may be explained in terms of Mott
hybridization\cite{Mott} of low-lying energy levels with their
Andreev-reflected images, in the spirit of our recent work.\cite{we12}
We leave this interesting question for future study.

Experimentally, a modification of the LDOS by one level
is usually difficult to observe (in particular, it is typically
not self-averaging). However similar effects may be of relevance
in more experimentally accessible geometries (e.g., in planar
NS junctions, where the corrections to the LDOS would be
self-averaging due to a large area of the NS interface).
Also, our calculation for the unitary symmetry class may
provide helpful intuition for treating systems without
TR symmetry breaking.

An important step in studying nonperturbative localization
effects in such proximity systems without TR symmetry
breaking was done in Ref.\ \onlinecite{TS2003}. There a
scattering-matrix formalism was used to find the modification
of the LDOS in TR-symmetric quantum wires close to the NS
interface, in comparison with the predictions from the
Usadel equations. It was found that the LDOS is reduced
within the energy window of the order $\Delta_\xi$ around
the Fermi level, which is qualitatively similar to our
results for wires with broken TR symmetry. It would be
interesting to extend the results of Ref.\ \onlinecite{TS2003}
to finite distances from the NS interface, possibly using
indications from our present work. Progress in this direction
may also be relevant for a theoretical analysis of
experiments on Josephson junctions via Anderson
insulators.\cite{Zvi}

Recently a great interest has been attracted to one-dimensional
hybrid structures hosting Majorana fermions at the NS boundary. Such
bound zero-energy eigenstates appear if the superconductor is of
a topological type.\cite{Oreg,Lutchyn} In the random-matrix context,
such systems are assigned to class B.\cite{Caselle,Bocquet,Ivanov}
An evidence in favor of Majorana excitations in an InSb nanowire
coupled to a superconductor has been reported in very recent experiments.
\cite{Mourik,Deng}
The zero-energy mode at the NS boundary is topologically
protected and hence robust with respect to disorder. It is this
robustness that makes the Majorana fermions promising candidates for
realization of fault-tolerant quantum computations\cite{Beenakker}
and motivates further intensive research in the field.

Disorder effects on the spectral and transport characteristics of
one-dimensional hybrid structures of class B models were studied both
analytically and numerically within the scattering matrix
formalism.\cite{Brouwer,Pientka,Pikulin}
The sigma-model approach developed in our
paper can also be applied to the hybrid wires of class B.
The characteristic distinction between classes D and B in the random-matrix
approach is that the two disconnected parts of the sigma-model manifold
contribute with either the same (class D) or opposite (class B) signs to
the partition function.\cite{Ivanov} In our one-dimensional problem,
this corresponds to the sign of the ``south-pole'' ($\lambda_F=-1$)
contribution in Eq.~(\ref{rho(x)-large-GA-paramag}).
By changing the difference
of the two wave functions in Eq.~(\ref{rho(x)-large-GA-paramag}) into their
sum, we can readily infer the LDOS in class B including the
spatial profile of the Majorana mode and the depletion of the local density
of other low-energy states. This calculation will be the subject of a
separate forthcoming publication.\cite{tobe}

\acknowledgments

Hospitality of EPFL where a significant part of this work has been done
is gratefully acknowledged.
This work was partially supported by the Russian Ministry
of Education and Science (Contract No. 8678) (Ya.F. and M.S.)
and the German Ministry of Education and Research (BMBF)
(P.O. and M.S.).


\appendix

\section{Sigma model for the NS junction in classes C and D}
\label{A:D}

In this Appendix we outline the derivation of the sigma model for a normal
metal -- superconductor junction both in the case of preserved (class C) and
violated (class D) spin symmetry. The form of the sigma-model action is known
to be universal and depends only on the symmetries of the underlying system but
not on its microscopic details. Using this fact, we will pick a particular
Hamiltonian which bears all necessary symmetry properties and facilitates the
derivation of the sigma model.

Our strategy is as follows. We assume that the dominant disorder in the sample
has the form of random potential. This potential disorder sets the elastic mean
free path $l$, which is the shortest scale in the problem, and preserves both
time-reversal and spin symmetry. This assumption allows us to consider both
superconducting and normal part of the junction on equal footing and describe
them by a single sigma model for a diffusive system. Then we introduce an
additional relatively weak source of disorder that breaks the TR symmetry
in the normal part of the junction. This is either random vector potential
(class C) or magnetic impurities (class D). Violation of the TR symmetry
occurs on a longer scale $L_c \gg l$. Retaining only massless degrees of
freedom, we finally obtain the sigma model applicable at distances longer than
$L_c$. This will be the unitary class sigma model with boundary conditions of
either class C or D.

The starting point of the derivation is the Bogolyubov--de Gennes Hamiltonian.
Its general form is given by Eq.\ (\ref{BdG}). The Hamiltonian is a $2 \times 2$
matrix in the Nambu space. We use the notation $\tau$ for Pauli matrices acting
in this space. Let us briefly discuss the structure of the Hamiltonian for our
particular model. The strongest disorder is the random potential $U$. Singling
out the corresponding term, we write the Hamiltonian as
\begin{equation}
 \hat{\mathcal{H}}_\text{BdG}
  = \hat{\mathcal{H}}_0 + \tau_z U\, .
 \label{H0U}
\end{equation}
The most general form of the term $\hat{\mathcal{H}}_0$ is
\begin{subequations}
\label{H0}
\begin{align}
 &\text{class C:} &
 \hat{\mathcal{H}}_0
  = \tau_z \xi(\mathbf{p}) + \tau_y \Delta
    - \mathbf{a}\cdot\mathbf{v}
  \, ,
\\
 &\text{class D:} &
 \hat{\mathcal{H}}_0
  = \tau_z \xi(\mathbf{p}) + \tau_y \Delta
    - \mathbf{b}\cdot\mathbf{s}
  \, .
\end{align}
\end{subequations}
Here we have included (i) the standard kinetic term, (ii) the superconducting
order parameter $\Delta$
(assumed to be real),
and (iii) the random vector potential $\mathbf{a}$ (for class C)
or the random exchange field $\mathbf{b}$ due to
magnetic impurities acting on electron spin (for class D).
In the superconducting part of the junction,
$\mathbf{a}=\mathbf{b}=0$, whereas in the normal wire, $\Delta=0$.
We proceed with the
derivation of the sigma model using the general form of the Bogolyubov--de
Gennes Hamiltonian and will specify the effect of particular terms later.

Local density of states is expressed in terms of the retarded Green function
(\ref{rho_Green}). This Green function can be generated from the path integral
with the action
\begin{equation}
 S
  = -i\int dx\, \Phi^+ (E + i0 - \hat{\mathcal{H}}_\text{BdG}) \Phi \, .
 \label{Sphi}
\end{equation}
Here the vector field $\Phi$ contains both commuting and Grassmann variables
and also has the structure in the Nambu space.

Bogolyubov--de Gennes Hamiltonian obeys the symmetry $\hat{\mathcal{H}} = -
\tau_y \Theta \hat{\mathcal{H}}^* \Theta^{-1} \tau_y$ that provides the onset
of specific soft modes -- superconducting cooperons. In order to include these
modes into the sigma model, we rewrite the action (\ref{Sphi}) in the doubled
form, introducing the particle-hole structure of the fields.
\begin{gather}
 S
  = -i\int dx\, \bar\Psi
    [(E + i0)\sigma_z \tau_z - \tau_z \hat{\mathcal{H}}_\text{BdG}]
    \Psi \, , \\
 \Psi
  = \frac{1}{\sqrt{2}} \begin{pmatrix}
      \Phi \\ i \tau_y \Theta \Phi^*
    \end{pmatrix} \, , \quad
 \bar\Psi
  = \frac{1}{\sqrt{2}} \begin{pmatrix}
      \Phi^\dagger \tau_z, & \Phi^T k \tau_x \Theta^{-1}
    \end{pmatrix}\, .
\end{gather}
We denote Pauli matrices acting in the particle-hole space by $\sigma$. Later,
we will see that this space corresponds to the retarded-advanced structure in
the Efetov parameterization of the unitary class sigma model.
The supersymmetry-breaking matrix $k=\diag(1,-1)_\text{FB}$.

The matrix $\Theta$ is either unity (when spin is preserved, class C) or $is_y$
(spin symmetry is violated, class D).
In both cases, the vectors $\Psi$ and
$\bar\Psi$ are related by the identity $\bar\Psi = (C \Psi)^T$ with the
charge-conjugation matrix $C$ dependent on the symmetry class:
\begin{subequations}
\begin{align}
 &\text{class C:} &
 C
  &= -\tau_x \begin{pmatrix}
       i \sigma_y & 0 \\
       0 & \sigma_x
     \end{pmatrix}_\text{FB}\, , \label{C-C}
\\
 &\text{class D:} &
 C
  &= i \tau_x s_y \begin{pmatrix}
       \sigma_x & 0 \\
       0 & i \sigma_y
     \end{pmatrix}_\text{FB}\, . \label{C-D}
\end{align}
\end{subequations}

Derivation of the sigma model proceeds with averaging the statistical weight
$e^{-S}$ over potential disorder. Assuming Gaussian white-noise statistics of
$U$, averaging yields a quartic term $\sim (\bar\Psi \Psi)^2$. This term is
further decoupled with the help of the auxiliary matrix field $\mathcal{Q}$. The
action acquires the form
\begin{equation}
 S
  = \int dx \left[
      \frac{\pi \nu_0}{8\tau}
      \mathop{\mathrm{str}} \mathcal{Q}^2 - i\bar\Psi \left(
        E\sigma_z \tau_z + \frac{i\mathcal{Q}}{2\tau}-\tau_z \hat{\mathcal{H}}_0
      \right) \Psi
    \right]\, .
 \label{SPsiQ}
\end{equation}
Here $\hat{\mathcal{H}}_0$ is defined in Eqs.\ (\ref{H0U}) and (\ref{H0}),
and $\nu_0$ is the density of states per one spin projection.
The field $\mathcal{Q}$ is normalized such that the infinitesimal imaginary
part $i0 \sigma_z \tau_z$ is replaced by $i \mathcal{Q}/2\tau$ where $\tau$ is
the disorder-induced mean free time.

Next, we integrate out the vector field $\Psi$ and perform the saddle-point
analysis of the resulting action
\begin{equation}
 S
  = \frac{\pi \nu_0}{8\tau} \int dx \mathop{\mathrm{str}} \mathcal{Q}^2
    -\frac{1}{2} \mathop{\mathrm{str}} \ln \left(
      E\sigma_z \tau_z + \frac{i\mathcal{Q}}{2\tau}-\tau_z \hat{\mathcal{H}}_0
    \right)\, .
 \label{SQ}
\end{equation}
The term proportional to $E$ as well as all the terms in $\hat{\mathcal{H}}_0$
other than the kinetic term [see Eq.\ (\ref{H0})] are relatively small. We
first consider saddle points of the action (\ref{SQ}) with $E = 0$ and
$\hat{\mathcal{H}}_0 = \tau_z \xi(\mathbf{p})$. There is always one saddle
point equivalent to the self-consistent Born approximation for the average Green
function. It corresponds to replacing $i0$ in the original action with
$i/2\tau$, hence $\mathcal{Q} = \sigma_z \tau_z$ is a legitimate saddle point.
We can generate other saddle points with the help of the global gauge symmetry
of the fermionic action (\ref{SPsiQ}). Indeed, let us change the variables
in Eq.~(\ref{SPsiQ}): $\Psi \mapsto U \Psi$, $\bar\Psi \mapsto \bar\Psi \bar U$
with some matrix $U$ obeying the relation $\bar U U = 1$. (Charge conjugation for
matrices is defined as $\bar U = C U^T C^T$.) This transformation cannot alter
the value of the action for $\mathcal{Q}$. At the same time, $U$ commutes with
$\xi(\mathbf{p})$ since $U$ is constant in space. Hence the rotation of $\Psi$
can be compensated by $\mathcal{Q} \mapsto U \mathcal{Q} U^{-1}$. This allows us
to generate the whole saddle manifold of the sigma model,
\begin{equation}
 \mathcal{Q}
  = U \sigma_z \tau_z \bar U \, .
\end{equation}
The manifold is characterized by the conditions $\mathcal{Q}^2 = 1$ and
$\mathcal{Q} = \bar{\mathcal{Q}} = C \mathcal{Q}^T C^T$
(this space of $Q$ matrices corresponds to
disordered systems of the symmetry class AI,
in the classification of Zirnbauer \cite{Koziy,zirnbauer:96}).

The sigma-model action is the result of the expansion of Eq.\ (\ref{SQ}) to the
first order in scalars $\Delta$ and $E$ and to the second order in vectors
$\nabla Q$, $\mathbf{a}$, and $\mathbf{b}$. Assuming $\mathbf{a}$ and
$\mathbf{b}$ are Gaussian white-noise random quantities, we obtain
\begin{multline}
  S
  = \frac{\pi\nu_0}{8} \int dx \mathop{\mathrm{str}} \Bigl\{
      D \big(\nabla \mathcal{Q} \big)^2
      + 4i \big( E \sigma_z \tau_z + i \tau_x \Delta \big) \mathcal{Q}
\\
      + L_\text{sb}
    \Bigr\} \, ,
\label{SQsigma}
\end{multline}
with the symmetry-breaking term
\begin{subequations}
\label{f-sb}
\begin{align}
\label{f-sb-C}
 &\text{class C:} &
  L_\text{sb}
  &=
  - D \langle\mathbf{a}\rangle^2 [\tau_z, \mathcal{Q}]^2
  \, ,
\\
\label{f-sb-D}
 &\text{class D:} &
  L_\text{sb}
  &=
  - \frac{\tau}{3} \langle\mathbf{b}\rangle^2 [\tau_z\mathbf{s}, \mathcal{Q}]^2
  \, .
\end{align}
\end{subequations}

The above action contains simultaneously all possible terms of our sigma model.
Now we analyze the theory in particular cases keeping only the relevant terms in
Eq.\ (\ref{SQsigma}). Let us start with the superconducting part of the
junction. Assuming a bulk superconductor with large density of states and
diffusion constant, we keep only $E$ and $\Delta$ terms and neglect fluctuations
of $\mathcal{Q}$. This allows us to fix $\mathcal{Q} = \mathcal{Q}_S$ at the
saddle point of the sigma-model action,
\begin{equation}
 \mathcal{Q}_S
  = \begin{pmatrix}
      \sigma_z \cos \theta_S & \sin \theta_S \\
      \sin \theta_S & -\sigma_z \cos \theta_S
    \end{pmatrix}_\text{N} \, ,
 \quad
 \tan \theta_S
  = \frac{i\Delta}{E} \, .
 \label{QS}
\end{equation}

Now consider the normal part of the junction first with preserved spin symmetry
(class C). Here $\Delta = 0$ while random magnetic field,
$\langle\mathbf{a}\rangle^2 \neq 0$, has a strong effect violating the TR
symmetry. This leads to the additional constraint $[\tau_z, \mathcal{Q}] = 0$
making the matrix $\mathcal{Q}$ diagonal in the Nambu space. The condition
$\mathcal{Q} = \bar{\mathcal{Q}}$ with $C$ from Eq.~(\ref{C-C}) leads to the
representation
\begin{equation}
  \mathcal{Q}
  = \begin{pmatrix}
      Q & 0 \\
      0 & \Xi Q^T \Xi^T
    \end{pmatrix}_\text{N} \, ,
\label{calQ}
\end{equation}
where $\Xi$ is the matrix defined in Eq.~(\ref{Xi-C}).
In the normal part of the junction, the sigma model action (\ref{SQsigma})
can be written in terms of $Q$ with the only constraint $Q^2 = 1$.
Then one obtains exactly the action (\ref{SD4}) (with $\nu=\nu_0$)
for the unitary class sigma model, with the particle-hole structure
playing the role of the retarded-advanced space.

Let us now consider the case of broken spin symmetry (class D). The matrix
$\mathcal{Q}$ has a structure in the additional spin space and obeys the
relation $\mathcal{Q} = \bar{\mathcal{Q}}$ with $C$ from Eq.\ (\ref{C-D}).
Taking into account random exchange field, $\langle\mathbf{b}\rangle^2 \neq 0$,
we get the constraint $[\tau_z \mathbf{s}, \mathcal{Q}] = 0$. It makes
$\mathcal{Q}$ diagonal in the Nambu space and trivial (unit matrix) in the spin
space. The representation (\ref{calQ}) again applies but now with $\Xi$ from
Eq.\ (\ref{Xi-D}). In terms of $Q$, the sigma model action is again given by
Eq.\ (\ref{SD4}).
The trace over the spin space doubles the density of states
in the prefactor of the action: $\nu=2\nu_0$.

The last ingredient of the theory is the boundary term (\ref{Sboundary4}). In
terms of the matrix $\mathcal{Q}$, boundary action is known to
be\cite{Efetov-book}
\begin{equation}
 S_\Gamma
  = -\frac{1}{2} \sum_i \mathop{\mathrm{str}} \ln \big(
      1 + e^{-2\beta_i} \mathcal{Q}_S \mathcal{Q}
    \big) \, .
\end{equation}
Using Eqs.\ (\ref{QS}) and (\ref{calQ}), we trace the Nambu (and optionally
spin) space and obtain the boundary action in the form
\begin{multline}
 S_\Gamma
  =
  - \frac{1}{2} \sum_i \mathop{\mathrm{str}} \ln \Big[
      1 + e^{-2\beta_i} \cos\theta_S (Q \sigma_z - \sigma_z \Xi Q^T \Xi^T) \\
      - e^{-4\beta_i} Q \Xi Q^T \Xi^T
    \Big]\, .
    \label{boundary-action-general}
\end{multline}
If the gap in the superconductor $\Delta$ is sufficiently large, we may
neglect $\cos\theta_S$ in Eq.~(\ref{boundary-action-general}) and obtain
the simplified boundary action (\ref{Sboundary4}). An accurate estimate
for this approximation gives the condition $\sum_i {\cal T}_i^{1/2} E \ll \Delta$
in class C and $\sum_i {\cal T}_i^{1/2} \tilde{\lambda}_B E \ll \Delta$
in class D, where ${\cal T}_i$
are Andreev transmission coefficients given by Eq.~(\ref{Andreev-transmission})
and  $\tilde{\lambda}_B$ is the characteristic
scale of $\lambda_B$ involved. Up to a factor of order 2, we may estimate
$\sum_i {\cal T}_i^{1/2} \sim \sum_i T_i = g_N$, the normal conductance of the
SN interface. We also estimate
$\tilde{\lambda}_B \sim \max(\delta,\Delta_\xi)/ E$, where $\delta$ is the
mean level spacing in the N part, as defined in Eq.~(\ref{delta-definition}).
As a result, we find the applicability conditions for the boundary term
(\ref{Sboundary4}):
\begin{equation}
\Delta \gg
\begin{cases}
g_N E & \text{in class C,} \\
g_N \max (\delta, \Delta_\xi) & \text{in class D.}
\end{cases}
\label{large-Delta}
\end{equation}

\section{Derivation of the LDOS expressions
(\ref{rho(x)C}) and (\ref{rho(x)D})}
\label{A:derivation}

\subsection{Efetov's parametrization}
\label{A:Q}

We adopt the standard Efetov's parametrization \cite{Efetov-book}
of the $4\times4$ matrix $Q$ for the unitary symmetry class:
\be
\label{Q4}
  Q
  =
  U_\eta^{-1}
  Q_0
  U_\eta
  \, ,
\ee
where the FB-diagonal central part, $Q_0$, contains only commuting variables:
\begin{subequations}
\label{Q0}
\begin{gather}
  Q_0 =
  \begin{pmatrix}
    Q_0^\text{FF} & 0 \\
    0 & Q_0^\text{BB}
  \end{pmatrix}_\text{FB}
  \, ,
\\
  Q_0^\text{FF}
  =
  \begin{pmatrix}
    \lambda_F & e^{i\vp_F}\sqrt{1-\lambda_F^2} \\
    e^{-i\vp_F}\sqrt{1-\lambda_F^2} & -\lambda_F
  \end{pmatrix}_\text{RA}
 \, ,
\\
  Q_0^\text{BB}
  =
  \begin{pmatrix}
    \lambda_B & ie^{i\vp_B}\sqrt{\lambda_B^2-1} \\
    ie^{-i\vp_B}\sqrt{\lambda_B^2-1} & -\lambda_B
  \end{pmatrix}_\text{RA}
 \, ,
\end{gather}
\end{subequations}
while Grassmann variables reside in the RA-diagonal matrix
\begin{subequations}

\begin{gather}
\label{Ueta}
  U_\eta =
  \begin{pmatrix}
    u & 0 \\
    0 & v
  \end{pmatrix}_\text{RA}
  \, ,
\\
  u =
  \exp
  \begin{pmatrix}
    0 & \sigma \\
    \sigma^* & 0 \\
  \end{pmatrix}_\text{FB}
  \, ,
\quad
  v =
  \exp
  \begin{pmatrix}
    0 & \rho \\
    \rho^* & 0 \\
  \end{pmatrix}_\text{FB}
  \, .
\end{gather}
\end{subequations}
Integration variables change in the intervals
$\lambda_F\in[-1,1]$,
$\lambda_B\in[1,\infty)$,
$\vp_{F,B}\in[0,2\pi)$,
with the invariant measure
\be
\label{dQ-measure}
  dQ
  =
  \frac{d\lambda_B d\lambda_F}{(\lambda_B-\lambda_F)^2}
  \frac{d\vp_B d\vp_F}{(2\pi)^2} \,
  d\rho^* d\rho \, d\sigma^* d\sigma \, .
\ee

\subsection{Boundary term}

Substituting Eqs.~(\ref{Q4}) and (\ref{Q0}) into the boundary
action (\ref{Sboundary4}), after some algebra
we obtain for the class C [$\Xi$ is given by Eq.~(\ref{Xi-C})]:
\be
\label{S-Gamma-C}
  S_\Gamma
  =
  \left[
    1
    -
    \frac{\lambda_B-\lambda_F}{2}
    (\sigma-\rho^*)(\sigma^*+\rho)
    \frac{\partial}{\partial\lambda_B}
  \right]
  S_0^\text{C}(\lambda_B) \, ,
\ee
where the action $S_0^\text{C}(\lambda_B)$ is given by Eq.~(\ref{SGamma0C}).

For class D [$\Xi$ is given by Eq.~(\ref{Xi-D})] we get a similar result:
\begin{multline}
\label{S-Gamma-D}
  S_\Gamma
  =
  \left[
    1
    -
    \frac{\lambda_B-\lambda_F}{2}
    (\sigma-\rho^*)(\sigma^*+\rho)
    \frac{\partial}{\partial\lambda_F}
  \right]
  S_0^\text{D}(\lambda_F) \, ,
\end{multline}
where the action $S_0^\text{D}(\lambda_F)$ is given by Eq.~(\ref{SGamma0D}).

\subsection{Expression for the LDOS in terms of the effective quantum mechanics}

In the language of the effective quantum mechanics \cite{EL83},
with the coordinate along the wire playing the role of imaginary time,
calculation of the functional integral (\ref{rho-def1}) for
$\corr{\rho_E(x)}$ is performed in two steps.
First we evaluate the functional integral
along the segment $[x,L]$ of the wire, which generates the ``singlet''
wave function
\be
  \Psi(Q(x);s)
  =
  e^{-2Hs} \circ 1
  \, ,
\ee
where $s=(L-x)/\xi$ is the dimensionless distance from the open end.
By symmetry, the ``radial'' wave function $\Psi(Q(x);s)$ depends only
on the variables $\lambda_F$ and $\lambda_B$.

At the point $x$, the LDOS preexponent $\str(k\Lambda Q(x))$
introduces the multiplet of states $Q\Psi(Q)$.
Then, evaluation of the functional integral at the segment $[0,x]$
amounts to following the evolution of the multiplet $P\equiv Q\Psi(Q)$
from the observation point $x$ down to the SN interface.
The resulting matrix $P$ is known\cite{EL83,Efetov-book}
to be of the form (\ref{Q0}) with
the elements modified according to
\begin{subequations}
\label{g-subs}
\begin{gather}
  \lambda_F\to g_F\lambda_F, \qquad
  \sqrt{1-\lambda_F^2}\to f_F\sqrt{1-\lambda_F^2} \, ,
\\
  \lambda_B\to g_B\lambda_B, \qquad
  \sqrt{\lambda_B^2-1}\to f_B\sqrt{\lambda_B^2-1} \, .
\end{gather}
\end{subequations}

The evolution equations for the functions $f_F$ and $f_B$ which determine
the density-density response function (RA-off-diagonal components)
have been derived in Ref.~\onlinecite{EL83}.
For the LDOS, we need the RA-diagonal components, $g_F$ and $g_B$,
whose evolution is described by the coupled differential equations
\begin{subequations}
\label{g-transfer}
\begin{align}
  \frac12
  \frac{\partial g_F}{\partial t}
   &= -Hg_F
     +\frac{1 - \lambda_F^2}{\lambda_F}\, \frac{\partial g_F}{\partial \lambda_F}
     +\frac{\lambda_B}{\lambda_F} M \Delta g \, ,
\label{gF}
\\
  \frac12
  \frac{\partial g_B}{\partial t}
   &= -Hg_B
     +\frac{\lambda_B^2 - 1}{\lambda_B}\, \frac{\partial g_B}{\partial \lambda_B}
     +\frac{\lambda_F}{\lambda_B} M \Delta g \, ,
\label{gB}
\end{align}
\end{subequations}
where $t$ is the distance from the point $x$ measured in units of $\xi$,
$\Delta g = g_B - g_F$, and
\be
  M
  =
  \frac{\lambda_F\lambda_B - 1}{(\lambda_B-\lambda_F)^2}
  \, .
\ee
The initial conditions for the evolution (\ref{g-transfer})
are $g_F = g_B = \Psi(Q(x);s)$ at $t = 0$.

Calculation of the functional integral (\ref{rho-def1})
along the segment $[0,x]$ is equivalent to so solving
Eqs.~(\ref{g-transfer}) up to $t=x/\xi$. As a result,
one ends up with a single integral over ${\cal Q} \equiv Q(0)$
at the SN interface:
\be
\label{rho-Q0}
  \corr{\rho_E(x)}
  =
  \Re
  \int d{\cal Q} \,
  \frac{\str(k\Lambda P({\cal Q};t,s))}{4} \, e^{-S_\Gamma[{\cal Q}]}
  \, .
\ee
Using Eqs.~(\ref{Q0}) and (\ref{g-subs}), we express
\be
  \frac{\str(k\Lambda P)}{4}
  =
  \Phi_+
  -
  (\sigma\sigma^* + \rho\rho^*)
  \Phi_-
\label{PhiPsi}
\ee
in terms of two functions:
\be
  \Phi_\pm = \frac{g_B\lambda_B \pm g_F\lambda_F}{2} \, .
\ee
Combining Eqs.~(\ref{gF}) and (\ref{gB}) one immediately obtains \cite{DM}
that the function $\Phi_-$ exhibits a simple evolution with the Hamiltonian
(\ref{H}):
\be
  \frac{\partial\Phi_-}{\partial t} = -2H \Phi_- \, .
\ee
Taking into account the boundary condition
$\Phi_-=(\lambda_B-\lambda_F)\Psi(Q(x),s)$ at $t=0$
and the definition (\ref{tilde-H}) of the Hamiltonian $\tilde H$,
one arrives at
\be
  \Phi_-({\cal Q};t,s)
  =
  (\lambda_B-\lambda_F) \Psi(\lambda_F,\lambda_B;t,s)
  \, ,
\ee
where
\be
  \Psi(\lambda_F,\lambda_B;t,s)
  =
  e^{-2\tilde H t} \Psi(Q(x),s)
  \, .
\ee
Evolution of the function $\Phi_+({\cal Q};t,s)$ is more complicated
and can be in principle deduced from Eqs.~(\ref{g-transfer}).

Evaluating the integral (\ref{rho-Q0}) using Eqs.~(\ref{dQ-measure}),
(\ref{S-Gamma-C}) and (\ref{PhiPsi}), we get the LDOS for the symmetry
class C:
\begin{multline}
\label{rho(x)-Phi}
  \corr{\rho_E(x)}
  =
  \Re \Phi_+(1,1;t,s)
  +
  \frac12 \Re
  \int_{-1}^1
  d\lambda_F
  \int_1^\infty
  d\lambda_B
\\ {}
  \times
  \frac{\partial e^{-S_0(\lambda_B)}}{\partial\lambda_B} \,
  \Psi(\lambda_F,\lambda_B;t,s) \, ,
\end{multline}
where the first term comes from the anomaly associated with the singularity
of the measure (\ref{dQ-measure}) at the origin.\cite{Efetov-book}
Though the general evolution of $\Phi_+$ described by Eqs.~(\ref{g-transfer})
is quite complicated, its value at the origin, $\Phi_+(\Lambda)$, can be
readily found. Indeed, in the absence of coupling to the superconductor
($T_i=0$), one should reproduce the normal-state result, $\corr{Q}=\Lambda$.
Hence $\Phi_+(1,1;t,s)=1$, leading to Eq.~(\ref{rho(x)C})\, .

Similar calculations for the symmetry class D [with Eq.~(\ref{S-Gamma-D})
instead of (\ref{S-Gamma-C})] lead to Eq.~(\ref{rho(x)D})\, .

\section{Evolution at the Mott scale: diffusion and drift}
\label{A:diff-drift}

In this Appendix, we approximately calculate the wave function
$\Psi(\lambda_F, \lambda_B; t,s)$ defined in
Eq.~(\ref{full-wave-function-definition}) for $t \gg 1$ and
$|\kappa|^2 \ll 1$. First consider the wave function
\begin{equation}
\Psi(\lambda_F, \lambda_B; 0,s) = e^{-2Hs} \circ 1 \, .
\label{B:wave-function1}
\end{equation}
It interpolates between $\Psi(\lambda_F,\lambda_B)\equiv 1$ at $s=0$
and $\Psi_0(\lambda_F,\lambda_B)$ given by Eq.~(\ref{Psi0}) at $s \to\infty$.
We will use the following properties of the wave function
(\ref{B:wave-function1}):
\begin{itemize}
\item[(1)]
It does not develop any singularity at $\lambda_B\in [1;\infty)$ and
$\lambda_F \in [-1;1]$.
\item[(2)]
It varies at the length scale $\lambda\sim |\kappa|^{-2} \gg 1$
in both $\lambda_F$ and $\lambda_B$. In particular, the dependence
on $\lambda_F$ is very weak.
\item[(3)]
It is universally normalized at the origin: $\Psi(1,1;0,s)\equiv 1$.
This follows from the supersymmetry (there is a zero mode of $H$
with $\Psi(1,1)=1$, and all the excited modes have $\Psi(1,1)=0$).
\end{itemize}

Then we note that the fermionic part $\tilde{H}_F$ of
the Hamiltonian (\ref{tilde-H}) has a discrete spectrum with a
level spacing of order one, and therefore, at $t \gg 1$,
only the lowest eigenstate of $\tilde{H}_F$ survives.
By the perturbation theory, its energy is of the order
$|\kappa|^4$, and, since we will be interested at the
logarithmic scales $t\sim \ln |\kappa|^{-2}$, it can
be neglected, as well as its $\lambda_F$ dependence.
Therefore, at $t\gg 1$, we can approximate
\begin{equation}
\Psi(\lambda_F, \lambda_B; t,s) \approx
e^{-2\tilde{H}_Bt} \Psi(1,\lambda_B; 0,s)\, .
\label{B:boson-evolution}
\end{equation}

To understand the evolution in $\lambda_B$, it is convenient
to change to the logarithmic coordinate $\theta_B$ defined by
\begin{equation}
\lambda_B=\cosh\theta_B .
\label{B:theta-definition}
\end{equation}
In this variable,  $\Psi(1,\theta_B; 0,s) \approx 1$ for
$\theta \lesssim \ln |\kappa|^{-2}$ (or even at larger $\theta_B$,
for small $s$). In terms of $\theta_B$, the bosonic Hamiltonian
takes the form
\begin{equation}
\tilde{H}_B = -\frac{1}{2} \frac{\partial^2}{\partial\theta_B^2}
-\frac{1}{2}\coth\theta_B \frac{\partial}{\partial\theta_B}
+\frac{\kappa^2}{16}\cosh\theta_B\, .
\label{B:hamiltonian-theta}
\end{equation}
This Hamiltonian consists of three terms. The first term
describes a diffusion in $\theta_B$. The second term produces
a drift towards the origin $\theta_B=0$. Note that the velocity of this
drift equals $1/2$ at $\theta_B\gg 1$ and differs from $1/2$ only near
the origin (in such a way that the total drift time changes by
a small correction of order one). The third, potential, term of the
Hamiltonian is relevant at $\theta_B \gtrsim \theta_M/2$, where we
define the Mott scale
\begin{equation}
\theta_M=2 \ln (|\kappa|^{-2}) \approx L_M/\xi
\label{B:Mott-theta}
\end{equation}
[we have introduced the coefficient 2 in the above definition in order
to relate $\theta_M$ to the conventional definition (\ref{LMott})
of the Mott length scale $L_M$].
The third term suppresses the wave function at $\theta_B > \theta_M/2$ for $t\gtrsim 1$,
regardless of the initial wave function (\ref{B:wave-function1}).
Therefore, at $t \sim 1$, the wave function
$\Psi(\lambda_F, \lambda_B; t,s)$ may be approximated as a
step function
\begin{equation}
\Psi(\theta_B)\approx\theta(\theta_M/2-\theta_B)\, ,
\label{B:step-function}
\end{equation}
independently of $\lambda_F$ and $s$. Note that the width of this step
function is of order one (in the units of $\theta_B$),
but, at our level of precision, we are not interested in details at
such a short length scale.

Finally, we may approximately calculate the result of the evolution
(\ref{B:boson-evolution}) by applying the first two terms of the Hamiltonian
(\ref{B:hamiltonian-theta}) to the step function (\ref{B:step-function}).
At our level of precision, we may also neglect the difference of the
drift velocity from $1/2$. We thus find the following result:
\begin{equation}
\Psi(\lambda_F, \lambda_B; t,s) \approx
\frac{1}{2}\left[ 1 - {\rm erf}\left(
\frac{\theta_B - \theta_M/2 + t}{2\sqrt{t}} \right) \right]\, ,
\end{equation}
where $\theta_B$ is related to $\lambda_B$ by Eq.~(\ref{B:theta-definition}).

In particular, at $\lambda_B=1$, we find
\begin{equation}
\Psi(\lambda_F, 1 ; t,s) \approx
\frac{1}{2}\left[ 1 - {\rm erf}\left(
\frac{t - \theta_M/2}{2\sqrt{t}} \right) \right]\, .
\label{B-erf-final}
\end{equation}
Note that the location of this erf-function is at $L_M/2$,
which is twice closer than the similar ``erf'' feature in the
correlation function of the local density of states in the
wire.\cite{GDP83,IOS2009}

The result (\ref{B-erf-final}) describes the crossover
of $\Psi(\lambda_F, 1 ; t,s)$ from one to zero around
$t\approx\theta_M/2$ and is valid only in the vicinity
of this point. To estimate the range of validity
of our approximations, we recall that we neglected
the details of the evolution of the initial state
at $t\lesssim 1$. This effectively introduces
an uncertainty $\Delta t \sim 1$ in the final
result (\ref{B-erf-final}). Its range of validity
may be therefore estimated from its insensitivity
to this uncertainty, which leads to the condition
$|t-\theta_M/2| \ll \theta_M$ (note that this range
of validity is larger than the width of the
crossover itself).


\section{Density of states at $E=0$ for class D}
\label{A:rho0(x)-D}

The LDOS in class D can be calculated exactly
for quasiparticles right at the Fermi energy ($E\to0$).
In this limit, the ``fermionic'' variable $q\in[0,\kappa]$
defined in Eq.~(\ref{pq}) is small, and the wave function
(\ref{full-wave-function-definition}) can be expanded as
\be
\label{Psi01}
  \Psi(p,q;t,s) = \psi_0(p;t,s) + q^2 \psi_1(p;t,s) + o(q^2) \, .
\ee
Substituting this series into the general expression
(\ref{rho(x)-large-GA-paramag}) we see that the smallness
of $\Psi(1,\lambda_B;t,s)-\Psi(-1,\lambda_B;t,s)\propto\kappa^2$
is compensated by the large value of the integral over $\lambda_B$,
leading to a finite limit at $\kappa\to0$:
\be
\label{rho0-psi1}
  \corr{\rho_0(x)}
  =
  1
  +
  2
  \int_0^\infty
  \psi_1(p;t,s)
  \, p \, dp
  \, .
\ee

The function $\Psi(p,q;t,s)$ is obtained through the Hamiltonian
evolution (\ref{full-wave-function-definition}).
Since the Hamiltonians $H$ and $\tilde H$ are related
by the transformation (\ref{tilde-H}), it suffices
to study only the $\tilde H$ evolution.
In the order $q^2$, the fermionic part (\ref{tilde-HF})
of the Hamiltonian $\tilde H$ acts as follows
(at $\kappa=0$):
\be
\nonumber
  \tilde H_F \begin{pmatrix} 1 \\ q^2 \end{pmatrix}
  =
  \begin{pmatrix} - q^2/8 \\ q^2 \end{pmatrix}
  + o(q^2) \, ,
\ee
and one immediately obtains the evolution of an arbitrary
function $\chi(p,q) = \chi_0(p) + q^2 \chi_1(p) + o(q^2)$
by the Hamiltonian $\tilde H$ to the desired accuracy:
\begin{multline}
\label{chi-solution}
  e^{-2\tilde H t} \chi(p,q)
  =
  \left(
    1
    +
    q^2 \frac{1-e^{-2t}}{8}
  \right)
  \hat U_t \chi_0(p)
\\{}
  +
  q^2 e^{-2t} \hat U_t \chi_1(p)
  +
  o(q^2)
  \, .
\end{multline}
Here
\be
  \hat U_t = e^{-2\tilde H_B^0 t}
\ee
is the evolution operator for the bosonic Hamiltonian (\ref{tilde-HB})
in the limit $\kappa=0$:\cite{IOS2009}
\be
\label{tilde-HB0}
  \tilde H_B^0
  =
  \frac{1}{8}
  \left(
  - \frac{1}{p} \frac{\partial}{\partial p} p^3 \frac{\partial}{\partial p}
  + p^2
  \right) \, .
\ee
This Hamiltonian defined at the semiaxis $p>0$ has a continuous
spectrum $E_k=(k^2+1)/8$ with the eigenfunctions $K_{ik}(p)/p$.

The general expression (\ref{full-wave-function-definition})
for $\Psi(p,q;t,s)$ can be identically rewritten in the form
\begin{multline}
  \Psi(p,q;t,s)
  =
  e^{-2\tilde H t} \Psi_0(p,q)
\\{}
  +
  e^{-2\tilde H t} (p^2-q^2) e^{-2\tilde H s} \frac{1-\Psi_0(p,q)}{p^2-q^2}
  \, ,
\end{multline}
where we used the relation (\ref{tilde-H}) and extracted the zero mode of $H$.
Now expanding in $q^2$ and successively applying Eq.~(\ref{chi-solution}),
we arrive at
\be
\label{psi1-infty}
  \psi_1(p;t,\infty)
  =
  \frac{1-e^{-2t}}{8} \hat U_t \Psi_0(p)
  + e^{-2t} \hat U_t \Pi_0(p) \, ,
\ee
\begin{multline}
\label{psi1-finite}
  \psi_1(p;t,s)
  =
  \psi_1(p;t,\infty)
  +
  \frac{1-e^{-2(t+s)}}{8} \hat U_t p^2 \hat U_s Y(p)
\\
  + e^{-2(t+s)} \hat U_t p^2 \hat U_s Z(p)
  - e^{-2t} \hat U_{t+s} Y(p)
  \, ,
\end{multline}
where
\begin{gather}
\nonumber
  \Psi_0(p) = \Psi_0(p,0)\, ,
\qquad
  \Pi_0(p) = \partial_{q^2}\Psi_0(p,q) \bigr|_{q=0}\, ,
\\[6pt]
  Y(p) = \frac{1-\Psi_0(p)}{p^2} \, ,
\qquad
  Z(p) = \frac{Y(p)-\Pi_0(p)}{p^2} \, .
\nonumber
\end{gather}

The evolution operator $U_t$ becomes trivial
in the basis of the eigenfunctions of the Hamiltonian (\ref{tilde-HB0}),
where it amounts to multiplication by $e^{-(k^2+1)t/4}$.
Transition to the basis of $K_{ik}(p)/p$ is known as
the Lebedev-Kontorovich transformation.\cite{Lebedev-Kontorovich}
For the zero mode $\Psi_0(p,q)$ given by Eq.~(\ref{Psi0}), we obtain
\begin{subequations}
\label{LK}
\begin{align}
\label{LK1}
&  \Psi_0(p)
  =
  \int_0^\infty k\, dk\,
  \frac{k^2+1}{2} \tanh\frac{\pi k}{2}
  \frac{K_{ik}(p)}{p} \, ,
\\
\label{LK2}
&  \Pi_0(p)
  =
  \int_0^\infty k\, dk\,
  \frac{k^2+5}{8} \tanh\frac{\pi k}{2}
  \frac{K_{ik}(p)}{p} \, ,
\\
\label{LK3}
&  Y(p)
  =
  \frac{2}{\pi}
  \int_0^\infty \frac{k\, dk}{k^2+1}
  \sinh\frac{\pi k}{2}
  \frac{K_{ik}(p)}{p} \, ,
\\
\label{LK4}
&  Z(p)
  =
  \frac{2}{\pi}
  \int_0^\infty \frac{k\, dk}{(k^2+1)(k^2+9)}
  \sinh\frac{\pi k}{2}
  \frac{K_{ik}(p)}{p} \, .
\end{align}
\end{subequations}
In order to evaluate the action of the composite operator
$\hat U_t p^2 \hat U_s$ in Eq.~(\ref{psi1-finite}) one has
to reexpand $pK_{ik}(p)$ in the eigenfunctions of the Hamiltonian
(\ref{tilde-HB0}) which is performed with the help of the
Lebedev-Kontorovich transformation:
\be
\label{<k|p2|l>}
  p K_{ik}(p)
  =
  \frac{1}{2}
  \int_0^\infty l\, dl\,
  \frac{(l^2-k^2) \sinh\pi l}{\cosh\pi l-\cosh\pi k}
  \frac{K_{il}(p)}{p} \, .
\ee

Substituting Eqs.~(\ref{LK1}) and (\ref{LK2})
into Eqs.~(\ref{rho0-psi1}) and (\ref{psi1-infty}),
and integrating over $p$ with the help of
\be
  \int_0^\infty K_{ik}(p) dp
  =
  \frac{\pi}{2\cosh\pi k/2} \, ,
\ee
we obtain the result (\ref{rho0(x)-D})
for the LDOS in the infinite wire, $\corr{\rho_0^{(\infty)}(x)}$.

The general expression for the zero-energy LDOS in a finite wire
can be written with the help of Eqs.~(\ref{rho0-psi1}),
(\ref{psi1-finite}), (\ref{LK3}), (\ref{LK4}) and (\ref{<k|p2|l>}) as

\begin{widetext}

\begin{multline}
\label{rho0-gen}
  \corr{\rho_0(x)}
  =
  \corr{\rho_0^{(\infty)}(x)}
  -
  2 e^{-2t}
  \int_0^\infty \frac{k\, dk}{k^2+1}
  \tanh\frac{\pi k}{2}
  e^{-(k^2+1)(t+s)/4}
\\{}
  +
  \frac{1}{4}
  \int_0^\infty
  \!\!
  \int_0^\infty
  \frac{k\, dk \, l\, dl\, (k^2-l^2)}{\cosh\pi k-\cosh\pi l}
  \sinh\frac{\pi k}{2}
  \sinh\frac{\pi l}{2}
  \left[
  \frac{1}{k^2+1}
  -
  \frac{e^{-2(t+s)}}{k^2+9}
  \right]
  e^{-(k^2+1)s/4 -(l^2+1)t/4}
  \, .
\end{multline}

\end{widetext}


\begin{thebibliography}{99}
\bibitem{Ehrenfest-comment}
Trajectory randomization is assumed to be fast.
The opposite case is considered in Ref.~\onlinecite{Taras-Semchuk-Altland}.
\bibitem{Delta-xi}
The same energy scale can be expressed as
$\Delta_\xi \sim (\nu A \xi)^{-1}$, where
$\nu$ is the bulk density of states and $A$
is the wire cross section. The two expressions agree
up to a numerical coefficient, due to the relation
(\ref{xi-def}) for the localization length.
\bibitem{com-crossover}
For moderate symmetry breaking, crossover between symmetry classes
has been studied in Ref.~\onlinecite{Koziy}.
\end{thebibliography}
\end{document}